\documentclass[aps,prd,twocolumn,showpacs,floats,floatfix,letterpaper,nofootinbib,superscriptaddress,]{revtex4}

\usepackage{amssymb,amsmath,latexsym,mathrsfs}
\usepackage{graphicx,subfigure}
\usepackage{epsfig}
\usepackage{varioref,xr-hyper}
\usepackage{color}
\usepackage{multirow}
\usepackage{array}

\newcommand{\nua}[1]{\ensuremath{\rlap
           {\kern-2.5pt\ensuremath
           {\overset{\scriptscriptstyle(-)}{\phantom{\nu}}}}
           {\ensuremath{{\nu}_{#1}}}}}

\begin{document}

\title{Testing 3+1 and 3+2 neutrino mass models with cosmology and short baseline experiments}
\thanks{Preprint number: DFTT 05/2012}

\author{Maria Archidiacono}
\affiliation{Physics Department, Universit\`a di Roma ``La Sapienza'' and INFN, P.le Aldo Moro 2, 00185, Rome, Italy}

\author{Nicolao Fornengo}
\affiliation{Department of Physics, University of Torino and INFN, Via P. Giuria 1, I--10125 Torino, Italy}

\author{Carlo Giunti}
\affiliation{Department of Physics, University of Torino and INFN, Via P. Giuria 1, I--10125 Torino, Italy}

\author{Alessandro Melchiorri}
\affiliation{Physics Department, Universit\`a di Roma ``La Sapienza'' and INFN, P.le Aldo Moro 2, 00185, Rome, Italy}

\begin{abstract}
Recent results from short--baseline neutrino oscillation experiments
and Cosmic Microwave Background anisotropy measurements
suggest the presence of additional sterile neutrinos.
In this paper we properly combine these data sets to
derive bounds on the sterile neutrino
masses in the 3+1 and 3+2 frameworks, finding
a potentially good agreement between the two datasets.
However, when galaxy clustering is included in the analysis
a tension between the oscillation and cosmological data is clearly present.
\end{abstract}

\maketitle

\section{Introduction}

In recent years, the impressive experimental discoveries in two fields of investigation,
namely neutrino physics and cosmic microwave background anisotropies, have revolutionized
our knowledge in particle physics and cosmology. 
Neutrino oscillations experiments have not only firmly established that neutrino are massive
and mixed particles
(for reviews, see e.g. Refs. \cite{Giunti:2007ry,Bilenky:2010zza,Xing:2011zza}),
but have also provided
precise measurements of the three-neutrino mixing parameters
(see the recent global fits in Refs. \cite{Tortola:2012te,Fogli:2012ua}).
On the other hand, the measurements of the angular spectrum of the
Cosmic Microwave Background (CMB) anisotropies
(see e.g. Ref. \cite{wmap7})  have not only fully confirmed the expectations of the standard
cosmological scenario but also provided a precise determination of most of its parameters.
Moreover, with the continuous experimental improvements, a clear interplay between neutrino
physics and cosmology is emerging.
Neutrinos are indeed a fundamental energy component in modern cosmology. 
A cosmological neutrino background is expected in the standard model and affects both 
the shape of the CMB and the formation of cosmological structures (see e.g. Ref. \cite{Lesgourgues:2006nd}). 
The recent cosmological data have provided a clear evidence 
(more than $5$ standard deviations) for the existence
of the primordial neutrino background  and have strongly constrained
the absolute neutrino mass scale
(see e.g. Ref. \cite{Hannestad:2007tu}).

However, the measurements of CMB anisotropies made by the ACT (Atacama Cosmology Telescope)
\cite{act} and SPT (South Pole Telescope) \cite{spt} experiments, when combined with the measurements of the Hubble
constant $H_0$ and galaxy clustering data, have provided interesting hints
for an {\it extra} relativistic weakly interacting component, coined {\it dark radiation}.
Parameterizing this energy component with the effective number of neutrino species
$N_{\rm eff}$, the recent data bound it to $N_{\rm eff}=4.08\pm0.8$ at $95 \%$ C.L. 
(see e.g. Ref. \cite{Hou:2011ec,Archidiacono:2011gq,zahn,Hamann:2011hu}) whereas
the standard prediction for only three active neutrino species is
$N_{\rm eff}=3.046$ \cite{Mangano:2005cc}. While this result should be taken with some grain of salt,
since it is derived from a combination of cosmological data and some tension does
exist between the data (see e.g. Ref. \cite{calaratra}) it is anyway interesting since a fourth, or
fifth, neutrino species seems also suggested by short--baseline (SBL)
oscillation experiments.
The appearance and disappearance data
of several SBL experiments can be explained by the mixing of the three active neutrinos
with one or two additional sterile neutrinos
in the so-called 3+1 and 3+2 models
(see Refs. \cite{Kopp:2011qd,Giunti:2011gz,Giunti:2011hn,Giunti:2011cp,Karagiorgi:2012kw,Donini:2012tt}).

This work is aimed to determine the masses of the sterile neutrinos
in 3+1 and 3+2 models
using data from SBL experiments and recent cosmological data
and check if the results are mutually compatible.
Finally, we combine the bounds from the two different analyses
to have a joint probability for the masses of sterile neutrinos.
The paper is organized as follows:
in Sec.~\ref{sec:SBL} and in Sec.~\ref{sec:ii}
we present the data sets we make use of, the method we adopt to analyze them
and the results we obtain
regarding the SBL experiments and in the cosmological context, respectively;
in Sec.~\ref{sec:iiii} the joint analysis method and results are shown;
finally we summarize our conclusions in Sec.~\ref{sec:iiiii}.

\section{Neutrino oscillations analysis}
\label{sec:SBL}

The short--baseline neutrino oscillation analysis is performed following
Refs.~\cite{Giunti:2011gz,Giunti:2011hn,Giunti:2011cp}.

We consider
3+1 and 3+2 neutrino spectra
in which
$\nu_{e}$,
$\nu_{\mu}$,
$\nu_{\tau}$
are mainly mixed with
$\nu_1$,
$\nu_2$,
$\nu_3$, whose masses are much smaller than 1 eV
and there are one or two additional massive neutrinos,
$\nu_4$ and
$\nu_5$,
which are mainly sterile
and have masses of the order of 1 eV.
Short-baseline oscillations
are generated by the large squared-mass differences
$\Delta{m}^2_{41}$
and
$\Delta{m}^2_{51}$,
with:
\begin{equation}
\Delta{m}^2_{51}\geq\Delta{m}^2_{41}\gg\Delta{m}^2_{31}\gg\Delta{m}^2_{21}
\,.
\label{hierarchy}
\end{equation}
The small squared--mass differences
$\Delta{m}^2_{21}$
and
$\Delta{m}^2_{31}$
which
generate, respectively,
solar and atmospheric neutrino oscillations
(see Refs.\cite{Giunti:2007ry,Bilenky:2010zza,Xing:2011zza})
have negligible effects in SBL oscillations
and are ignored in the following.
The two heavy neutrino masses
$m_{4}$ and $m_{5}$
which are probed by cosmological data
are simply connected to the squared--mass differences
relevant for SBL oscillations
by:
\begin{equation}
m_{4} \simeq \sqrt{\Delta{m}^2_{41}}
\,,
\qquad
m_{5} \simeq \sqrt{\Delta{m}^2_{51}}
\,.
\label{masses}
\end{equation}

\begin{table}[t]
\begin{center}
\begin{tabular}{|c|c|c|}
\hline
&
{\bf 3+1}
&
{\bf 3+2}
\\
\hline
$\chi^2_{\text{min}}$			& $142.1	$ & $134.1	$ \\
$\text{NDF}$				& $138		$ & $134	$ \\
$\text{GoF}$				& $39\%		$ & $48\%	$ \\
$\Delta{m}^2_{41}\,[\text{eV}^2]$	& $1.62		$ & $0.89	$ \\
$|U_{e4}|^2$				& $0.035	$ & $0.018	$ \\
$|U_{\mu4}|^2$				& $0.0086	$ & $0.015	$ \\
$\Delta{m}^2_{51}\,[\text{eV}^2]$	& $		$ & $1.61	$ \\
$|U_{e5}|^2$				& $		$ & $0.022	$ \\
$|U_{\mu5}|^2$				& $		$ & $0.0047	$ \\
$\eta$					& $		$ & $1.57\pi	$ \\
\hline
\end{tabular}
\end{center}
\caption{ \label{tab:sbl}
Values of
$\chi^{2}_{\text{min}}$,
number of degrees of freedom (NDF),
goodness--of--fit (GoF)
and
best--fit values of the mixing parameters
obtained in our 3+1 and 3+2 fits of short--baseline oscillation data.
}
\end{table}

We fit the data set of short-baseline neutrino oscillation experiments
corresponding to the GLO--HIG analysis in Ref.~\cite{Giunti:2011cp},
in which the low-energy MiniBooNE neutrino \cite{AguilarArevalo:2008rc} and antineutrino 
\cite{AguilarArevalo:2010wv,Zimmerman:2011hy,Djurcic:2012jf}
data corresponding to the so-called "MiniBooNE low--energy anomaly" are not considered,
since they induce a strong tension between appearance and disappearance data
(see the discussions in Refs.~\cite{Giunti:2011hn,Giunti:2011cp}).
We made the following two improvements with respect to the analysis presented in Ref.~\cite{Giunti:2011cp}:
\begin{enumerate}
\item
We used the reactor neutrino fluxes presented in the recent White Paper on
light sterile neutrinos
\cite{Abazajian:2012ys},
which update Refs.~\cite{Mueller:2011nm,Huber:2011wv}.
The new fluxes are about 1.3\% larger than those we used before,
which were taken from the reactor antineutrino anomaly publication
\cite{Mention:2011rk}.
\item
We replaced the KamLAND bound on $|U_{e4}|^2$
with a more powerful constraint obtained from solar neutrino data
\cite{Giunti:2009xz,Palazzo:2011rj,Palazzo:2012yf}.
Taking into account the recent measurement of
$|U_{e3}|^2$
in the Daya Bay
\cite{An:2012eh}
and RENO
\cite{Ahn:2012nd}
reactor neutrino experiments
($
|U_{e3}|^2
=
\sin^{2}\vartheta_{13}
=
0.025 \pm 0.004
$),
from Fig.~1 of Ref.~\cite{Palazzo:2012yf}
we inferred the approximate upper bound
$
|U_{e4}|^2
=
\sin^{2}\vartheta_{14}
\lesssim
0.02
$
at $1\sigma$
(see Ref.~\cite{Giunti-NUTURN-2012}).
\end{enumerate}

In our analysis of SBL neutrino oscillation data we apply first the standard $\chi^2$ method.
The minimum value of $\chi^2$,
the number of degrees of freedom,
the goodness--of--fit
and the corresponding
best--fit values of the oscillation parameters are presented in Tab.~\ref{tab:sbl}.
The results concerning the 3+1 and 3+2 fits
are similar to those reported, respectively, in Ref.~\cite{Giunti:2011cp} for the GLO--HIG case
and Ref.~\cite{Giunti:2011gz},
with small variations due to the consideration of different data sets.
From Tab.~\ref{tab:sbl}
we can see that in both the 3+1 and 3+2 frameworks the global
goodness--of--fit is satisfactory.

\begin{figure}[t]
\includegraphics[bb=7 22 577 565, scale=0.4]{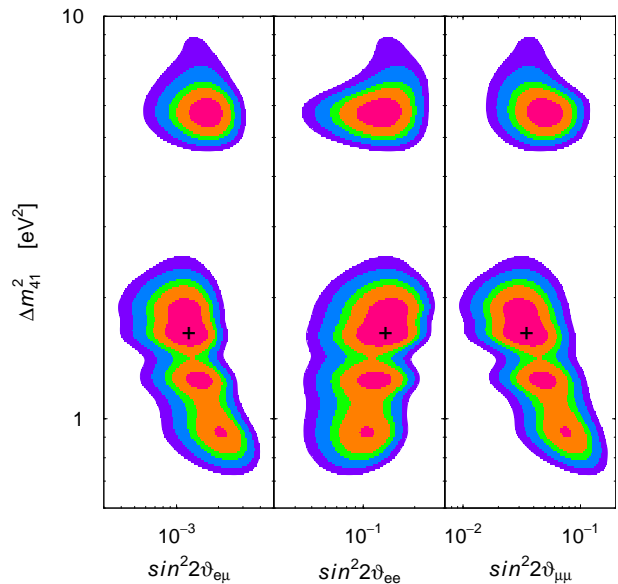}
\caption{Allowed regions in the
$\sin^{2}2\vartheta_{e\mu}$--$\Delta{m}^2_{41}$,
$\sin^{2}2\vartheta_{ee}$--$\Delta{m}^2_{41}$ and
$\sin^{2}2\vartheta_{\mu\mu}$--$\Delta{m}^2_{41}$
planes
obtained from the global fit of short--baseline neutrino oscillation data in the 3+1 scheme
using the standard $\chi^2$ method. The best-fit point is indicated by a cross
(see Table.~\ref{tab:sbl}).
}
\label{sbl-3p1}
\end{figure}

The allowed regions of
$\Delta{m}^2_{41}$
versus the effective SBL oscillation amplitudes
$\sin^{2}2\vartheta_{e\mu}$,
$\sin^{2}2\vartheta_{ee}$ and
$\sin^{2}2\vartheta_{\mu\mu}$
(with
$\sin^{2}2\vartheta_{\alpha\beta}=4|U_{\alpha4}|^2|U_{\beta4}|^2$)
are shown in Fig.~\ref{sbl-3p1}.
These regions are relevant, respectively, for
$\nua{\mu}\leftrightarrows\nua{e}$,
$\nua{e}\to\nua{e}$ and
$\nua{\mu}\to\nua{\mu}$
oscillation experiments.
They are more similar to those shown in Fig.~3 of Ref.~\cite{Giunti:2011cp}
than the region presented in Ref.~\cite{Giunti-NUTURN-2012},
because the larger reactor antineutrino fluxes used in this analysis
increase the reactor antineutrino anomaly,
leading to a larger value of
$|U_{e4}|^2$,
which tends to cancel
the effect of the solar neutrino constraint.

\begin{figure}[t]
\includegraphics[bb=7 22 576 566, scale=0.4]{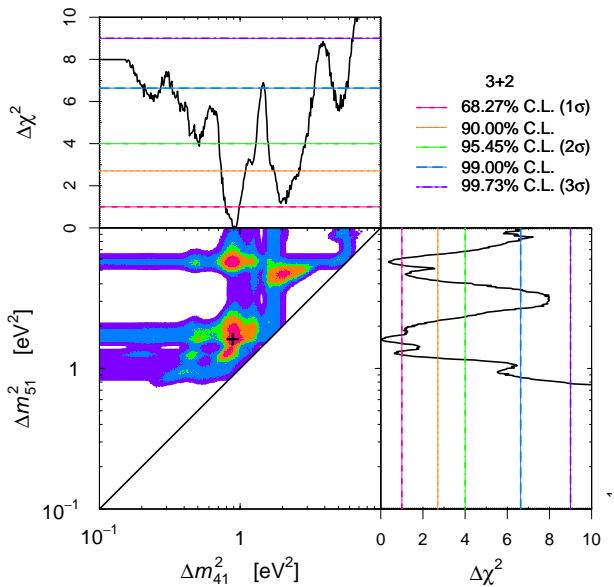}
\caption{Allowed regions in the $\Delta m_{41}^2$-$\Delta m_{51}^2$ plane and corresponding marginal
$\Delta\chi^2$'s obtained from the global fit of short--baseline neutrino oscillation data in 3+2 schemes
using the standard $\chi^2$ method. The best-fit point is indicated by a cross
(see Table.~\ref{tab:sbl}).
}
\label{sbl-3p2}
\end{figure}

The allowed regions in the
$\Delta{m}^2_{41}$-$\Delta{m}^2_{51}$
plane obtained in the 3+2 analysis
are shown in Fig.~\ref{sbl-3p2}.
One can see that the allowed regions are similar to those presented in Fig.~9 of Ref.~\cite{Giunti:2011gz},
with small variations due to the different considered data sets.

Since we want to perform a combined analysis of
SBL oscillation data and cosmological data
and the cosmological analysis is performed with the Bayesian method,
we have also analyzed the SBL oscillation data
with a Bayesian approach.
We assumed the sampling distribution of the data $D$:
\begin{equation}
p(D|\theta_{M},M)
\propto
e^{-\chi^2(D,\theta_{M})/2}
\,,
\label{sampling}
\end{equation}
where $M$ is the model
($M=3+1$
or
$M=3+2$),
$\theta_{M}$
is the corresponding set of oscillation parameters
(listed in Tab.~\ref{tab:sbl})
and
$\chi^2(D,\theta_{M})$
is the corresponding $\chi^2$ function.
The sampling probability is called ``likelihood'' when considered as a function of the parameters of the model.
In each of the two models,
we calculated the posterior probability distribution of the oscillation parameters
using Bayes' theorem:
\begin{equation}
p(\theta_{M}|D,M)
=
\frac{p(D|\theta_{M},M) p(\theta_{M}|M)}{p(D|M)}
\,,
\label{bayes}
\end{equation}
where $p(D|M)$ is easily calculated as a normalization constant.
We assumed a flat prior distribution in the logarithmic space of the oscillation parameters,
except for the CP--violating phase $\eta$
in the 3+2 spectrum (see Ref.~\cite{Giunti:2011gz})
for which we used a linear scale in the interval $[0,2\pi]$.
For
$\log(\Delta{m}^2_{41}/\text{eV}^2)$
and
$\log(\Delta{m}^2_{51}/\text{eV}^2)$
we considered the range $[-1,1]$.
For
$\log|U_{e4}|^2$,
$\log|U_{\mu4}|^2$,
$\log|U_{e5}|^2$,
$\log|U_{\mu5}|^2$
we considered the range $[-4,0]$.

Since we are interested in combining the results of the analysis of
SBL oscillation data with that of the cosmological data,
where the only shared parameters are the neutrino masses in Eq.(\ref{masses}),
we calculated the marginal posterior probability distributions of
the squared--mass differences by integrating
the posterior probability distribution
over the other oscillation parameters taking into account the scale of the flat prior.
For example, in the 3+1 model:
\begin{align}
\null & \null
p(\log\Delta{m}^2_{41}|D,3+1)
=
\int
d\log|U_{e4}|^2
\,
d\log|U_{\mu4}|^2
\nonumber
\\
\null & \null
\times
p(\log(\Delta{m}^2_{41}),\log|U_{e4}|^2,\log|U_{\mu4}|^2|D,3+1)
\,.
\label{marginal}
\end{align}
In this way,
we obtained
the posterior probability distribution of
$\Delta{m}^2_{41}$
in the 3+1 spectrum plotted in Fig.~\ref{3e1-a}
(thick green line exhibiting several sharp peaks)
and
the allowed regions in the
$\Delta{m}^2_{41}$--$\Delta{m}^2_{51}$
of the 3+2 spectrum
shown in Fig.~\ref{sbl2}.
Comparing with Fig.~\ref{sbl-3p2},
one can see that the Bayesian allowed regions are wider than those obtained with
the $\chi^2$ method.
The difference is due to the different method of marginalization with respect to the other mixing parameters
(mixing angles and CP--violating phase):
in the $\chi^2$ method one considers only the minimum of the $\chi^2$ in the range of each marginalized parameter,
whereas in the Bayesian method one must integrate the posterior probability density
over the marginalized parameter space.
Since the data do not constrain much the values of the marginalized parameters
(see Figs.~10--12 of Ref.~\cite{Giunti:2011gz}),
the Bayesian integration gives significantly different results from the $\chi^2$ marginalization.
The allowed vertical bands with constant value of
$\Delta{m}^2_{41}$
are due to the fact that one can have a comparable fit
for any value of $\Delta{m}^2_{51}$
and negligible $|U_{e5}|$ and $|U_{\mu5}|$,
which is effectively equivalent to a 3+1 framework.
The same applies to the allowed horizontal bands with constant value of
$\Delta{m}^2_{51}$.

\section{Cosmological analysis}
\label{sec:ii}

The cosmological analysis is performed in two different steps: first by analyzing CMB--only
data and then by further adding data from large scale structure and priors on the Hubble
parameter. The CMB analysis is performed by employing the following datasets: WMAP7 \cite{wmap7}, ACT \cite{act} and SPT \cite{spt}. The large scale structure analysis makes use of
information on dark matter clustering from the matter power spectrum
extracted from the SDSS--DR7 luminous red galaxy sample
\cite{red}. Finally, the Hubble parameter prior we use is based
on the latest Hubble Space Telescope observations \cite{hst}.

We analyze datasets up to $\ell_{\rm max}=3000$.
The analysis method we adopt is based on the publicly available Markov Chain Monte Carlo
(MCMC) package \texttt{CosmoMC} \cite{Lewis:2002ah} with a convergence
diagnostic done through the Gelman and Rubin statistic.

We sample the following six--dimensional standard set of cosmological parameters,
adopting flat priors on them: the baryon and cold dark matter densities
$\Omega_{\rm b} h^2$ and $\Omega_{\rm c} h^2$, the ratio of the sound horizon to the angular
diameter distance at decoupling $\theta$, the optical depth to reionization $\tau$,
the scalar spectral index $n_S$ and the overall normalization of the spectrum $A_S$.
We account for foregrounds contributions including three extra amplitudes:
the SZ amplitude, the amplitude of clustered point sources,
and the amplitude of Poisson distributed point--sources.
We consider purely adiabatic initial conditions and we impose spatial flatness.
In this work both active and sterile neutrinos are assumed to be fully
thermalized (for the non thermal case see, e.g., Ref. \cite{Dodelson:2005tp}).

The aim of this paper is to specifically test 3+1 and 3+2 neutrino mass models, by means
of a joint analysis of both cosmological and SBL experiments data. Therefore,
contrary to the typical approach (see e.g. Ref. \cite{Hamann:2010bk,Giusarma:2011ex,Archidiacono:2012gv}),
in the cosmological analysis we do not let the effective number of relativistic degrees of freedom $N_{\rm eff}$ to vary as a free parameter, instead we fix it at the values
$N_{\rm eff} = 3+1$ or $N_{\rm eff} = 3+2$ for the 3+1 and 3+2 schemes, respectively. This is
consistent with the assumptions done in the oscillation analysis and with the hypothesis of
cosmological full thermalization of all neutrino states (including the sterile ones;
see the recent discussions in Refs. \cite{Hannestad:2012ky,Mirizzi:2012we}).
Consistently to the analysis of Section \ref{sec:SBL}, we
fix the three active neutrinos to be massless and we allow the sterile neutrinos to have masses which vary as additional free parameters.
Since we are interested to sample the joint sensitivity of cosmological
and SBL neutrino data on the sterile--neutrinos mass parameters, in the cosmological
analysis we do not employ the neutrino mass fraction $f_{\nu}$ (as it is usually done), but
instead we sample directly $\log \Delta m_{41}^2$ and $\log \Delta m_{51}^2$. This implies a flat prior on those parameters.

\begin{table*}[t!]\footnotesize
\begin{center}
\begin{tabular}{|l|c|c|c|c|}
\hline
~ & ~ & ~ & ~ & ~ \\
~ & ~~{\bf 3+1} CMB only~~ & ~~{\bf 3+2} CMB only~~ & ~~{\bf 3+1} CMB+SDSS+HST~~ & ~~{\bf 3+2} CMB+SDSS+HST~~ \\
~ & ~ & ~ & ~ & ~ \\
\hline
~ & ~ & ~ & ~ & ~ \\
$\Omega_b h^2$ & $0.0224\pm0.0004$ & $0.0226\pm0.0004$ & $0.0224\pm0.0004$ & $0.0226\pm0.0004$\\
$\Omega_c h^2$ & $0.135\pm0.007$   & $0.156\pm0.009$   & $0.133\pm0.004$   & $0.156\pm0.004$\\
$\tau$         & $0.085\pm0.014$   & $0.087\pm0.015$   & $0.084\pm0.014$   & $0.086\pm0.014$\\
$H_0$          & $71.5\pm3.6$      & $73.6\pm4.4$      & $73.1\pm1.6$      & $74.6\pm2.0$\\
$n_s$          & $0.970\pm0.015$   & $0.985\pm0.016$   & $0.977\pm0.010$   & $0.990\pm0.010$\\
$\log(10^{10} A_s)$ & $3.21\pm0.05$ & $3.20\pm0.05$     & $3.19\pm0.04$     & $3.19\pm0.04$\\
~ & ~ & ~ & ~ & ~ \\
\hline
~ & ~ & ~ & ~ & ~ \\
$\Sigma m$ ($eV$)   & $<2.88$      & $<2.48$           & $<0.73$           & $0.58^{+0.12 \,\,(+0.45)}_{-0.13\,\,(-0.42)}$\\
~ & ~ & ~ & ~ & ~ \\
\hline
~ & ~ & ~ & ~ & ~ \\
$\chi^2_{\rm min}$ & $7529.5$      & $7532.2$          & $7578.5$          & $7581.1$\\
~ & ~ & ~ & ~ & ~ \\
\hline
\end{tabular}
\caption{
MCMC estimation of the cosmological parameters from the analysis of CMB--only data and from CMB data plus matter power spectrum information (SDSS) and a prior on $H_0$ (HST),
in the case of three massless active neutrinos and one massive sterile neutrino (3+1 scheme)
and assuming 3 massless active neutrinos plus 2 massive sterile neutrinos (3+2 scheme).
Neutrino mass upper bounds are reported at the $95 \%$ C.L., unless for the 3+2 CMB+SDSS+HST case
where we quote the best--fit value together with the $68 \%$ ($95 \%$)  C.L. interval.}
\label{cmb}
\end{center}
\end{table*}

\begin{figure*}[ht]
\includegraphics[scale=0.29]{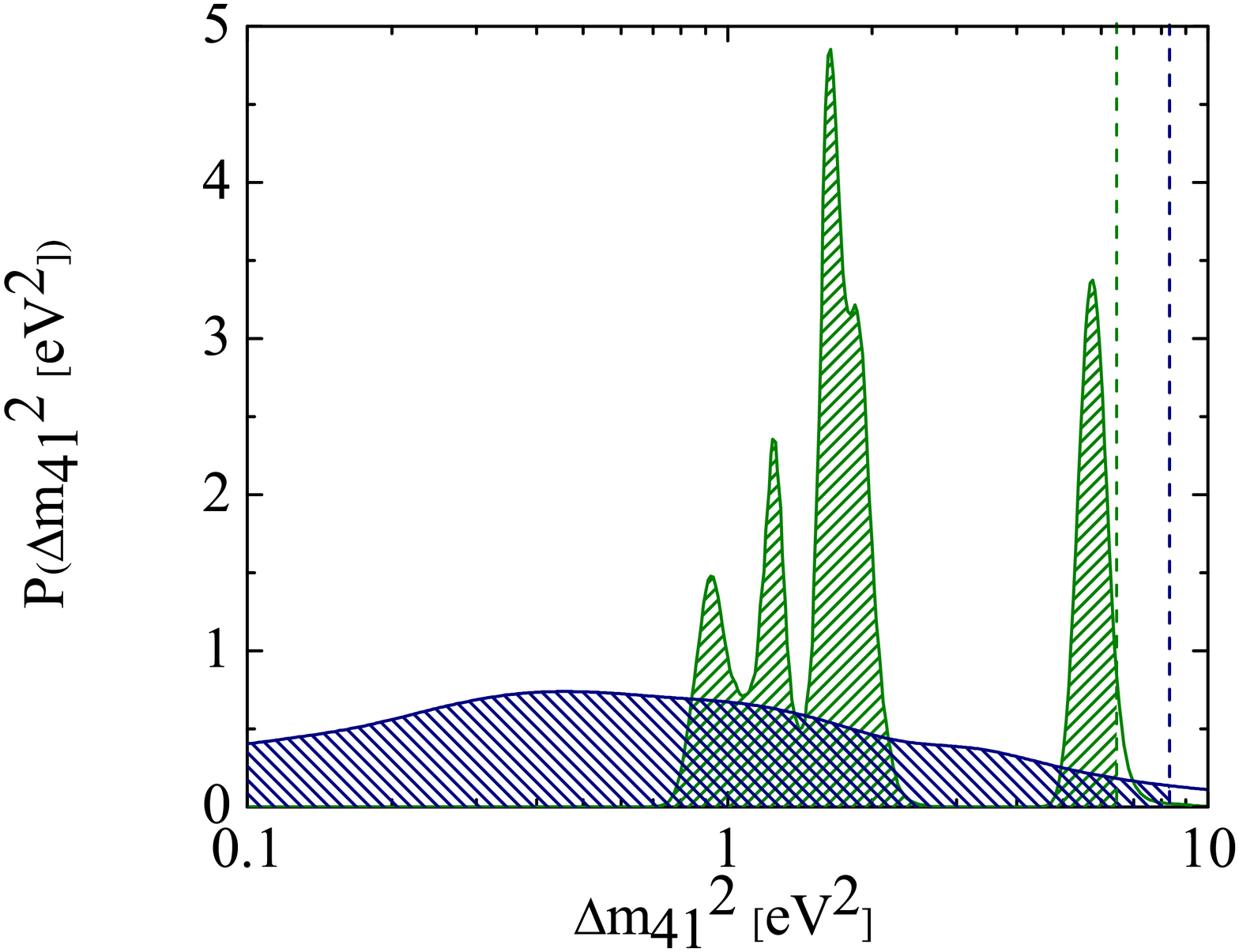}
\includegraphics[scale=0.29]{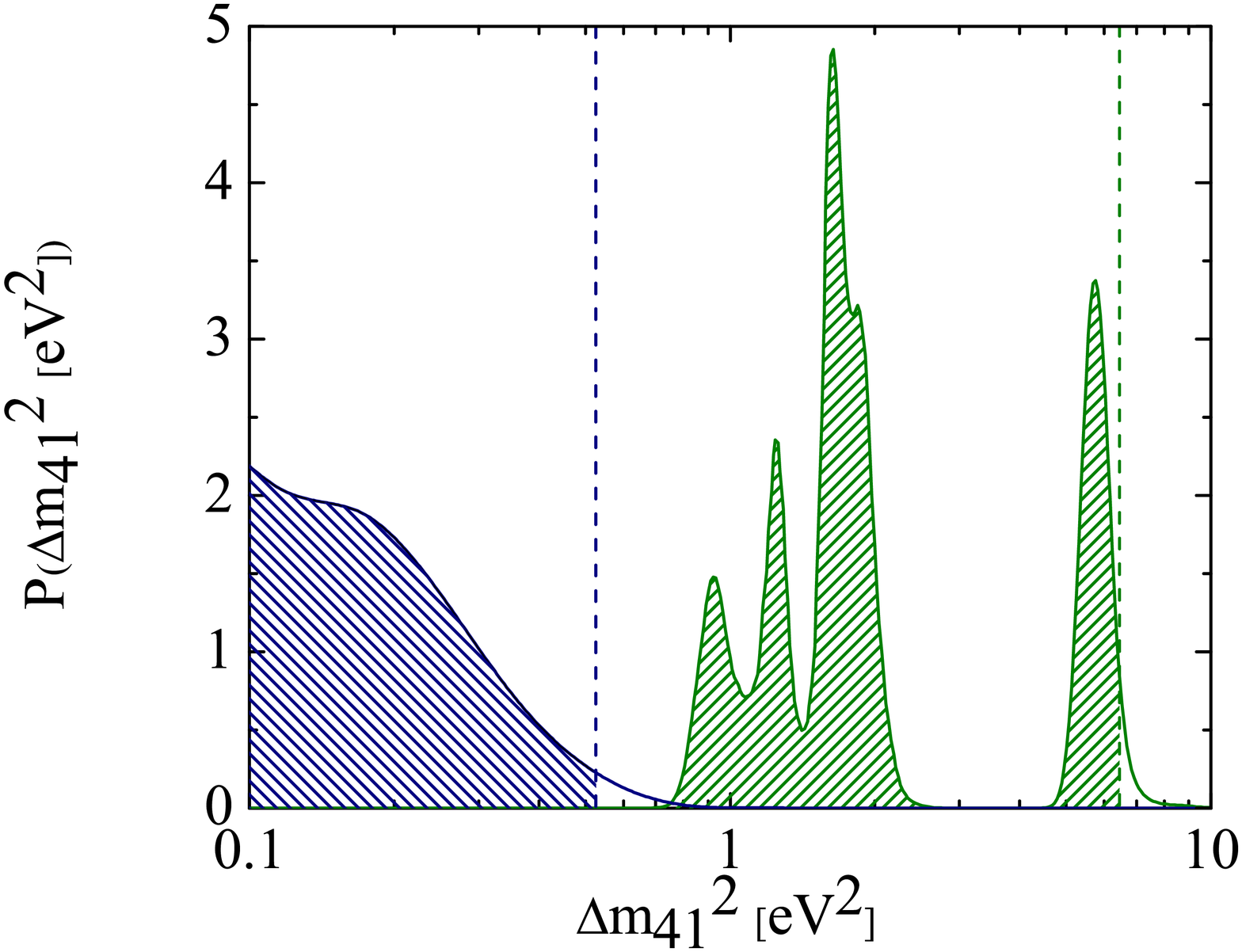}
\caption{
Marginal posterior probabilities obtained with a Bayesian analysis
for $\Delta m_{41}^2$ in the 3+1 scheme. The thick [green] solid line exhibiting several sharp peaks 
(the same in the two panels) refers to
the analysis of the short--baseline oscillation data alone. 
The blue line exhibiting a broad peak stands for the analysis of the cosmological data alone: CMB-only data for the left panel, CMB data implemented with SDSS and HST information for the
right panel. In all cases, the shaded regions
refer to the 95\% coverage of the probability distribution.}
\label{3e1-a}
\end{figure*}

\begin{figure*}[ht]
\includegraphics[scale=0.5]{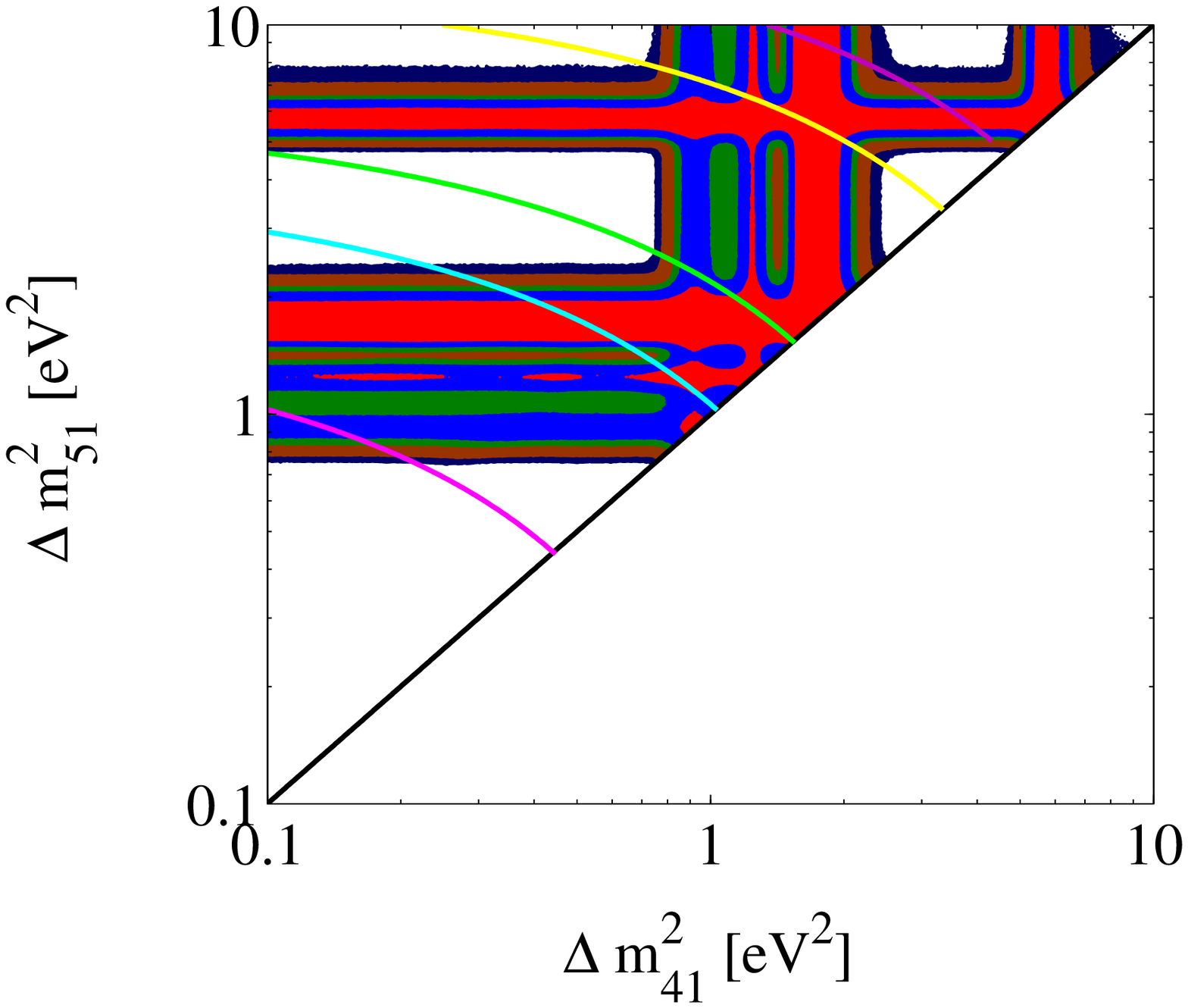}
\includegraphics[scale=0.5]{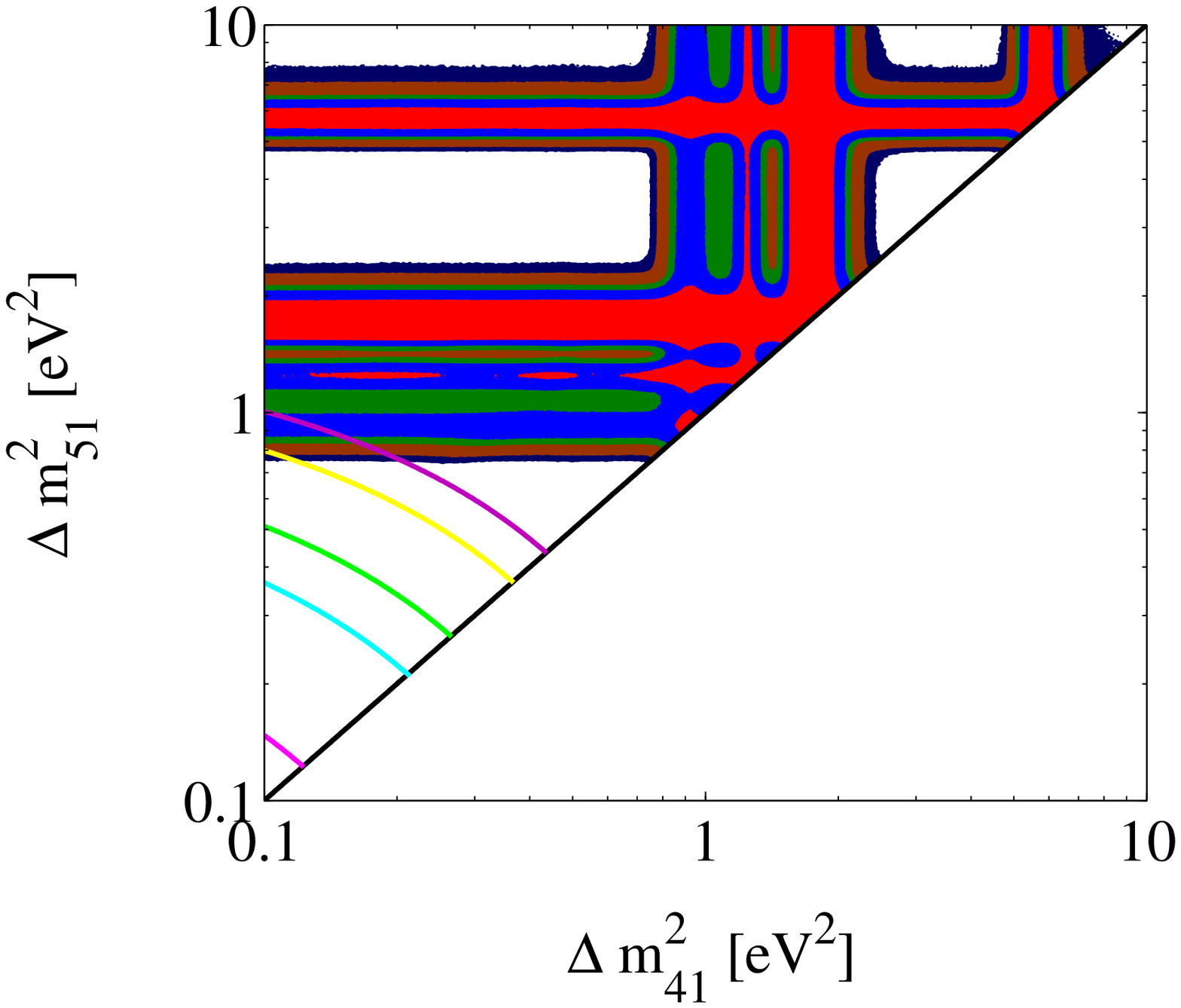}
\caption{Allowed regions in the $\Delta m_{41}^2$-$\Delta m_{51}^2$ plane obtained with a Bayesian analysis. The ``boxy'' regions (the same in the two panels) refer to the global analysis of the short--baseline oscillation data and are relative to the following confidence levels (from the innermost to the outermost region): 68.27\% (red), 90.00\% (light blue), 95.45\% (green), 99.00\% (brown) and 99.73\% (dark blue). The arc--shaped solid lines refer to the analysis of the cosmological data: the left panel stands for the CMB--only
dataset, while the right panel refers to the inclusion of the SDSS information and HST prior to the CMB
data. The different lines refer to the following confidence levels (from the lower curve to the upper curve, in each panel): 68.27\% , 90.00\%, 95.45\%, 99.00\% and 99.73\% .}
\label{sbl2}
\end{figure*}

Before attempting a joint analysis with the SBL data, which have been presented in the previous Section,
we report in
Tab.~\ref{cmb} the constraints on the cosmological parameters
using CMB--only data and CMB data plus SDSS information together with the HST prior,
and assuming: a 3+1 model with three massless active neutrinos and one massive sterile neutrino; a 3+2 model with three massless active neutrinos plus two massive sterile neutrinos. The
95\% C.L. mass bounds on the sterile neutrinos is 2.88 eV for the 3+1 scheme, while for
the 3+2 model the bound on the sum of the masses of the two additional
sterile neutrinos is 2.48 eV,
both of them share a 2$\sigma$ upper limit of about 1.24 eV,
when CMB--only data are used. These bounds drastically improve when also SDSS data and the HST prior are included in the analysis
(see Ref. \cite{GonzalezGarcia:2010un}),
reaching the value of 0.73 eV for the 3+1 case and about 1 eV for
the 3+2 case. Both the 3+1 and 3+2 schemes are statistically well acceptable, with no
noticeable preference in the minimal $\chi^2$. The only visible (and expected) difference between the
3+1 and 3+2 schemes is that 2 additional neutrinos require a
larger value of the dark matter abundance $\Omega_c h^2$, to compensate a delay of the equivalence time, which would instead be induced by the presence of an additional light degree of freedom  in the 3+2 case \cite{Hamann:2011ge}.
The correction due to non degeneracy between the mass of the first and the second sterile neutrino
in the 3+2 model
is of the order of precision of present numerical codes
and so undetectable
using only the present cosmological data (CMB and matter power spectra).
Moreover the degeneracies with other cosmological parameters
makes the detection of the neutrino mass differences impossible at the state of art (see Ref. \cite{Slosar:2006xb}).

Fig. \ref{3e1-a} shows the marginal posterior probability of the cosmological Bayesian analysis
for the 3+1 case, compared with
the results of the SBL study. The blue line exhibiting a broad peak stands for the analysis of the cosmological data alone and the left panel refers
to CMB-only data, while the right panel refers to the CMB data implemented with SDSS and HST information. The two panels of the figure show how the inclusion of SDSS and HST information is
relevant to set the more stringent constraint on the cosmological upper bound on the neutrino mass. The shaded regions refer to the 95\% C.L. coverage of the probability
distribution, from which the bounds on $m_4$ of Table \ref{cmb} are derived.
When compared with the SBL analysis and its 95\% C.L. mass intervals 
(three slightly discontinued ranges in the interval $0.93\,{\rm eV} < m_4 < 1.45\,\rm eV$ and a higher mass range $2.29\,{\rm eV} < m_4 < 2.59\,\rm eV$),
with a best fit at $m_4 = 1.27\,\rm eV$,
we notice that  CMB--only and SBL oscillation
data  are well compatible among them, with a significant overlap of the corresponding 95\% C.L.
regions. The 95\%  C.L. cosmological upper bound $m_4 < 2.88\, \rm eV$ disfavors the higher mass
SBL solution, while is perfectly compatible with the lower SBL mass ranges.
The combination of the cosmological and SBL datasets will therefore produce a clean allowed interval,
as shown in the next Section. Instead, when SDSS and HST information are included in the analysis,
SBL oscillations and cosmological data are in tension, with no overlapping 95\% C.L.

The analysis for the 3+2 scheme is shown in Fig. \ref{sbl2}, where C.L. regions in the
$\Delta m_{41}^2$--$\Delta m_{51}^2$ plane are reported. 
The SBL allowed regions clearly show a preference for at least a non--zero neutrino mass
($m_5$ with our choice of hierarchy in neutrino masses) and a global preference for $m_4 = 0.95\,\rm eV$ and $m_5 = 1.27\,\rm eV$. The cosmological data instead provide upper limits on both sterile neutrino masses,
with no clear preference for non--zero values. CMB--only data (left panel)  are well compatible 
with SBL results, with the 95\% C.L. upper bound
of the cosmological analysis consistent with the corresponding 95\% C.L.  regions of the
SBL analysis and its global best--fit point ($m_4 = 0.95\,\rm eV$ and $m_5 = 1.27\,\rm eV$). 
Also in the 3+2 case, the inclusion of SDSS and HST data produces tension between
SBL and cosmological analyses, as is manifest in the right panel of Fig. \ref{sbl2}, where only
a partial overlap at the 3$\sigma$ C.L. is present. Fig. \ref{sbl2} clearly
shows that the whole set of  cosmological
data will be instrumental in significantly reducing the degeneracy of the allowed solutions of the SBL analysis
when the joint analysis will be attempted in the next Section.

\section{Combined analysis}
\label{sec:iiii}

\begin{figure*}[ht]
\includegraphics[scale=0.5]{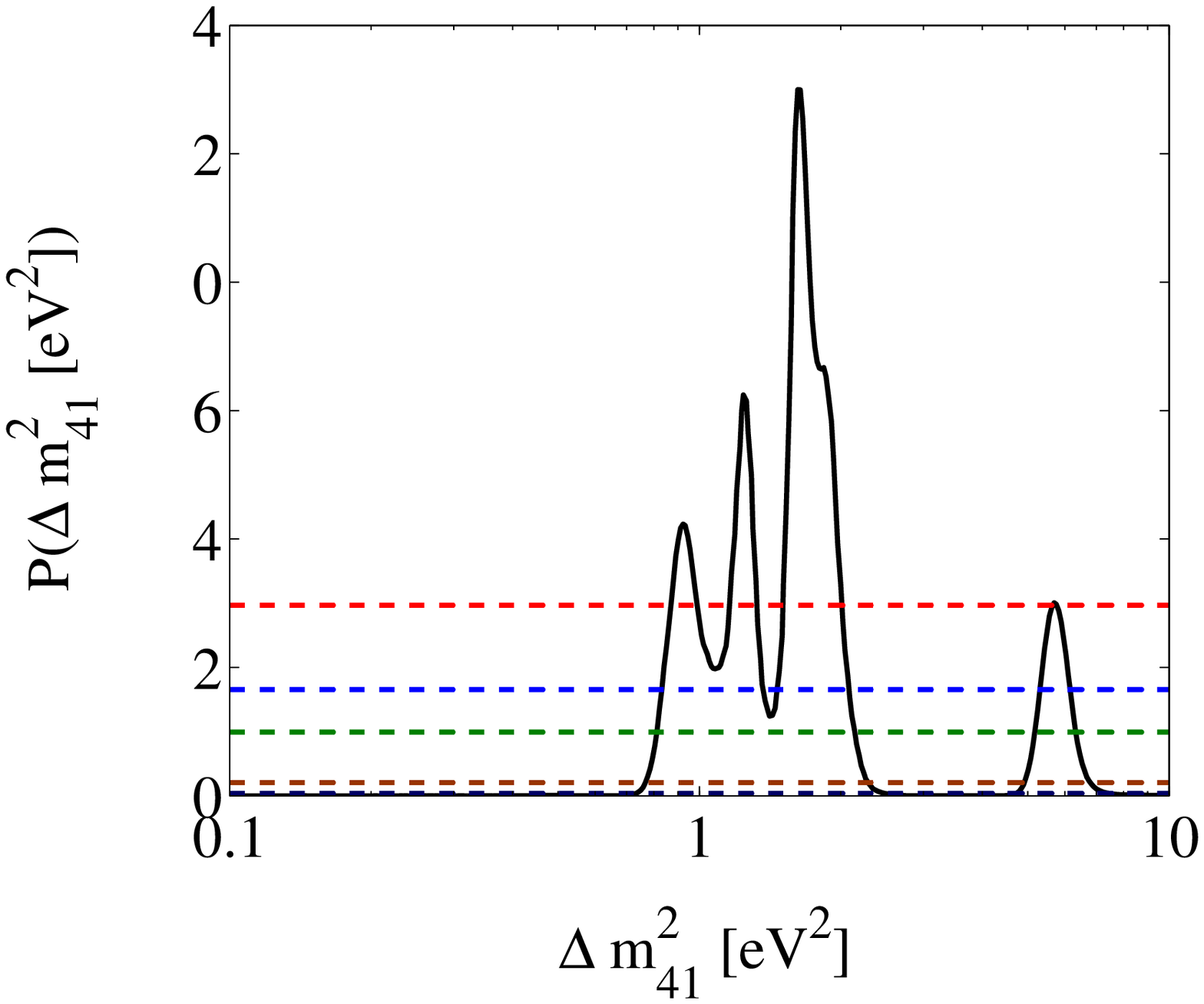}
\includegraphics[scale=0.5]{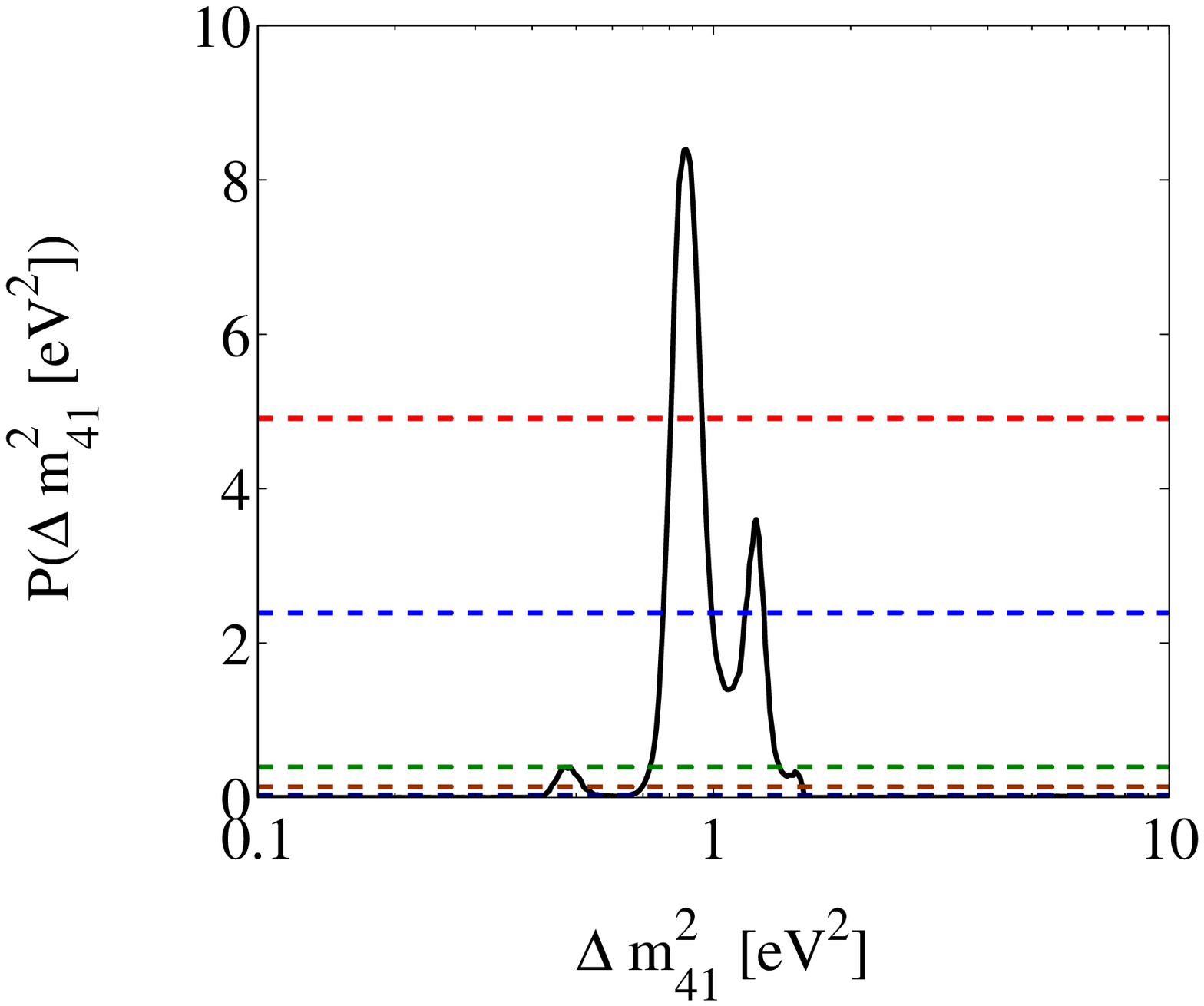}
\caption{
Marginal posterior probabilities obtained with a Bayesian analysis
for $\Delta m_{41}^2$ in the 3+1 scheme, for the {\bf joint} analysis of cosmological and short--baseline
data. Left panel: the cosmological analysis employs CMB--only data. Right panel: the cosmological analysis adds SDSS and HST information to the CMB data.
The horizontal dashed lines identify (from the lower curve to the upper curve, in each panel)
the credible intervals at 68.27\%, 90.00\%, 95.45\%, 99.00\% and 99.73\% C.L.}
\label{3e1-b}
\end{figure*}

\begin{figure*}[ht]
\includegraphics[scale=0.5]{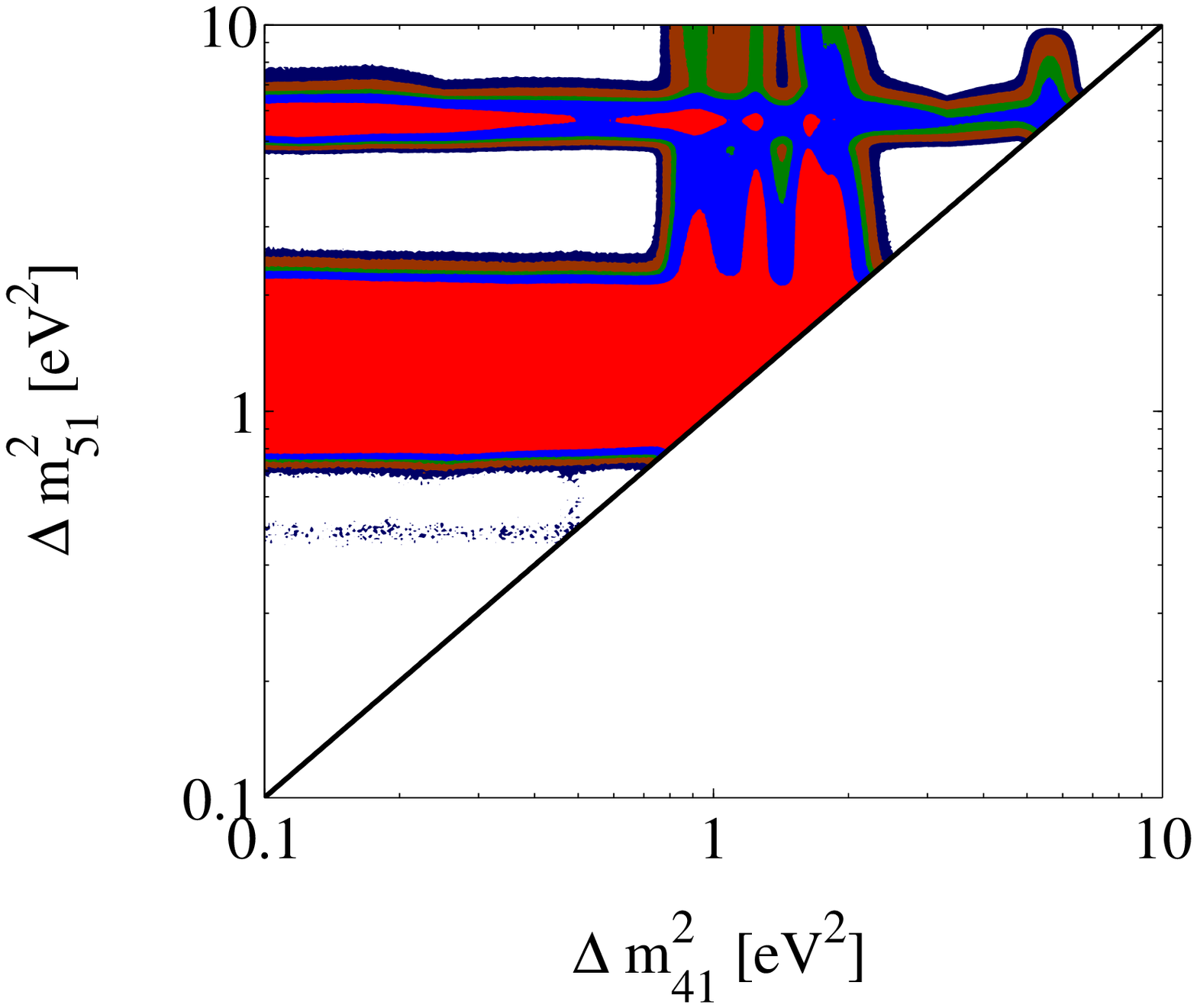}
\includegraphics[scale=0.5]{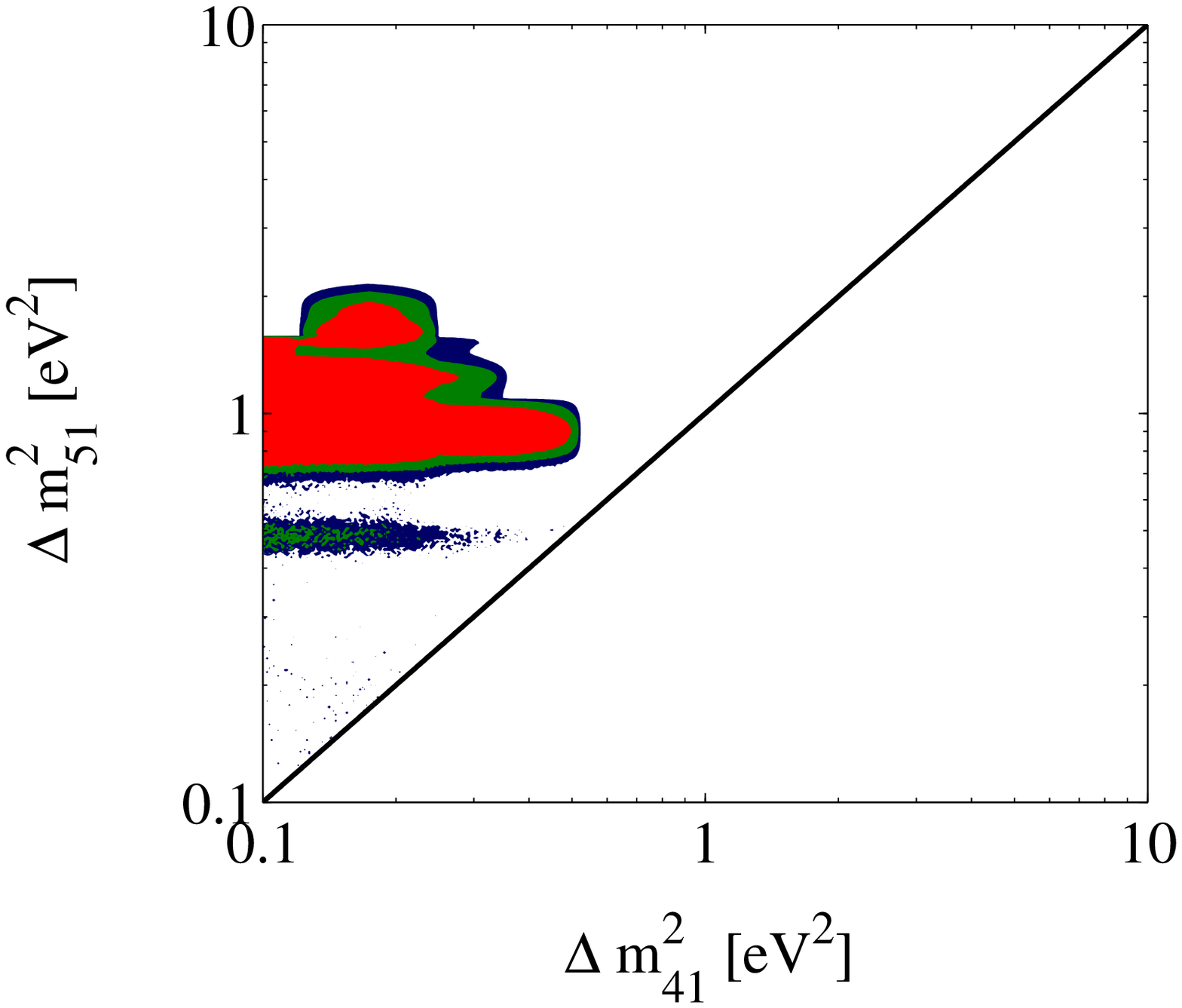}
\caption{Allowed regions in the $\Delta m_{41}^2$--$\Delta m_{51}^2$ plane obtained with a Bayesian approach, for the {\bf joint} analysis of  short--baseline and cosmological data. The different regions
(as in Fig. \ref{sbl2}) refer to the following confidence levels (from the innermost to the outermost region): 68.27\% (red), 90.00\% (light blue), 95.45\% (green), 99.00\% (brown) and 99.73\% (dark blue).  Left panel: SBL data plus the CMB--only dataset. Right panel: SBL data plus CMB, SDSS and HST data; in this case only 68.27\% (red), 95.45\% (green) and 99.73\% (dark blue) C.L. are reported.
}
\label{3e2}
\end{figure*}

The combined analysis of the SBL oscillation data and the cosmological observations has been
performed by merging the corresponding posterior probabilities.
Since the only relevant parameters
common to both sectors are the sterile neutrino masses $m_{4} \simeq \sqrt{\Delta{m}^2_{41}}$
and $m_{5} \simeq \sqrt{\Delta{m}^2_{51}}$ we can define a marginal posterior probability for the
joint analysis by directly multiplying  the SBL and cosmological marginal posterior probabilities relative
to the parameter of interest. For example, in the 3+1 case,
denoting by $D_{\text{C}}$ and $D_{\text{S}}$ the cosmological and SBL data
we have\footnote{
Since we assumed a flat prior for
$\theta=\log\Delta{m}^2_{41}$
in both the SBL and cosmological analyses,
using Bayes' theorem (\ref{bayes}) we have
$
p(\theta|D_{\text{C}+\text{S}})
\propto
p(D_{\text{C}+\text{S}}|\theta)
=
p(D_{\text{C}}|\theta)
p(D_{\text{S}}|\theta)
\propto
p(\theta|D_{\text{C}})
p(\theta|D_{\text{S}}).
$
}:
\begin{align}
\null & \null
p(\log\Delta{m}^2_{41}|D_{\text{C}+\text{S}},3+1) \propto \\
\null & \null
p(\log\Delta{m}^2_{41}|D_{\text{C}},3+1) \, \times \,
p(\log\Delta{m}^2_{41}|D_{\text{S}},3+1) \,,  \nonumber
\label{combined}
\end{align}
where the SBL probability is the one defined in Eq. (\ref{marginal}) and the 
cosmological probability is the one used in the analysis of the previous
section and obtained through \texttt{CosmoMC}.

The combined analysis for the 3+1 scheme is shown in Fig. \ref{3e1-b}. As usual, the left
panels refers to the case of CMB--only data in the cosmological sector, while the right panel
adds SDSS and HST datasets. The horizontal dashed lines identify the credible intervals at 68.27\%, 90.00\%, 95.45\%, 99.00\% and 99.73\% C.L. In the case of CMB--only data,  the
inclusion of the cosmological information to the SBL analysis disfavors the higher mass
SBL solution around 2.4 eV but maintains
the lower mass 95\% C.L. allowed intervals ($0.90\,{\rm eV} < m_4 <
1.46\,\rm eV$) and ($2.27\,{\rm eV} < m_4 <
2.51\,\rm eV$)
and the best--fit solution ($m_4 = 1.27\,\rm eV$).
When SDSS and HST information is added to the analysis, the allowed interval of the global analysis
shifts down to lower values of the sterile neutrino mass, due to the more stringent bound from the
cosmological sector. The 95\% C.L. mass range becomes
$0.85\,{\rm eV} < m_4 < 1.18\,\rm eV$,
and the best fit shifts down to $m_4 = 0.93\,\rm eV$.

The combined analysis for the 3+2 scheme is shown in Fig. \ref{3e2}, again for the case of CMB--only
data (left panel) and for the further inclusion of SDSS and HST data (right panel). The global results
are that at least one sterile neutrino needs to be massive, with a mass of the order of 1 eV ($m_5$ with our choice of hierarchy), while the second sterile neutrino can be massless. The marginalized
95\% intervals for the two neutrino masses are:
$m_4 < 2.51\,\rm eV$ and $0.86\,{\rm eV} < m_5 < 3.16\,\rm eV$ when
CMB--only data are considered;
$m_4 < 0.70\,\rm eV$ and $0.67\,{\rm eV} < m_5 < 1.35\,\rm eV$
for the full analysis which includes also SDSS and HST.

\section{Conclusions}
\label{sec:iiiii}

Measuring the number and the mass of sterile neutrinos is one of the most interesting challenges both in cosmology and in neutrino physics.
The existing cosmological data indicate that the energy density of the Universe may
contain dark radiation composed of one or two sterile neutrinos, which may correspond to those in
3+1 or 3+2 models which have been invoked for the explanation of
short--baseline neutrino oscillation anomalies.
We have performed analyses of the cosmological and SBL data in the frameworks of both the 3+1 and 3+2 models.
Then we have compared the results obtained with the same Bayesian method,
to figure out if the indications of cosmological and SBL data are compatible.

At the state of art, cosmological data are sensitive to the sum of neutrino masses, for which they give an upper limit at the scale of about 1 eV.
Hence they do not allow us to resolve the degeneracy between the mass of the first and the second sterile neutrino in a 3+2 model,
although in the numerical calculation we leave them as independent parameters.
Instead, short--baseline neutrino oscillations have a completely different parameterization and
in the 3+2 model the degeneracy between the two square mass differences $\Delta{m}^2_{41}$ and $\Delta{m}^2_{51}$ is broken.

The results of our analysis show that
the cosmological and SBL data give compatible results when the cosmological analysis takes into account only CMB data.
But if the information on the matter power spectrum coming from galaxies surveys are also considered
there is a tension between the sterile neutrino masses needed to have SBL neutrino oscillations
and the cosmological upper limit on the sum of the masses.

The combined analysis of cosmological and SBL data gives an allowed region
for $m_4$ in the 3+1 scheme around 1 eV.
In the 3+2 scheme, the cosmological data reduce the allowance of the second massive sterile neutrino
given by SBL data,
leading to a combined fit which prefers the case of only one massive sterile neutrino
at the scale of about 1 eV.

In conclusion, our analysis shows that cosmological data are marginally compatible
with the existence of one massive sterile neutrino with a mass of about 1 eV,
which can explain the anomalies observed in SBL neutrino oscillation experiments.
The case of massive sterile neutrinos is less tolerated by cosmological data
and in any case the second sterile neutrino must have a mass smaller than about
0.6 eV.

\begin{acknowledgements} 
NF and CG acknowledge the PRIN 2008 research grant
``Matter--Antimatter Asymmetry, Dark Matter and Dark Energy in the LHC Era"
(MIUR contract number: PRIN 2008NR3EBK) funded jointly by Ministero
dell'Istruzione, dell'Universit\`a e della Ricerca (MIUR), by
Universit\`a di Torino and by Istituto Nazionale di Fisica Nucleare.
We acknowledge INFN research grant FA51).
NF acknowledges support of the spanish MICINN
Consolider Ingenio 2010 Programme under grant MULTIDARK CSD2009- 00064 (MICINN).
AM work is supported by PRIN-INAF ``Astronomy probes fundamental physics".
MA acknowledges the European ITN project Invisibles (FP7-PEOPLE-2011-ITN, PITN-GA-2011-289442-INVISIBLES).
MA thanks the Department of Physics at University of Turin for hospitality
while this research was conducted.
\end{acknowledgements}


\begin{thebibliography}{99}

\bibitem{Giunti:2007ry} 
  C.~Giunti and C.~W.~Kim,
  ``Fundamentals of Neutrino Physics and Astrophysics'',
  Oxford University Press (Oxford, UK), 2007 (ISBN 978-0-19-850871-7)

\bibitem{Bilenky:2010zza} 
  S.~Bilenky,
  ``Introduction to the physics of massive and mixed neutrinos'',
  Lecture Notes in Physics {\bf 817}, 1 (2010), Springer (ISBN 978-3-642-14042-6)

\bibitem{Xing:2011zza} 
  Z.~Xing and S.~Zhou,
  ``Neutrinos in particle physics, astronomy and cosmology'',
  Zhejiang University Press, 2011 (ISBN 978-7-308-08024-8)

\bibitem{Tortola:2012te} 
  D.~V.~Forero, M.~Tortola and J.~W.~F.~Valle,
  arXiv:1205.4018 [hep-ph].

\bibitem{Fogli:2012ua} 
  G.~L.~Fogli, E.~Lisi, A.~Marrone, D.~Montanino, A.~Palazzo and A.~M.~Rotunno,
  arXiv:1205.5254 [hep-ph].

\bibitem{wmap7}
  E.~Komatsu {\it et al.},
  arXiv:1001.4538 [astro-ph.CO].

\bibitem{Lesgourgues:2006nd}
  J.~Lesgourgues and S.~Pastor,
  Phys.\ Rept.\  {\bf 429}, 307 (2006)
  [arXiv:astro-ph/0603494].

\bibitem{Hannestad:2007tu}
S.~Hannestad,
arXiv:0710.1952 [hep-ph].

\bibitem{act}
  J.~Dunkley {\it et al.},
  Astrophys.\ J.\ {\bf 739}, 52 (2011)
  [arXiv\-:1009.0866 [astro-ph.CO]].

\bibitem{spt}
  R.~Keisler {\it et al.},
  Astrophys.\ J.\ {\bf 743}, 28 (2011)
  [arXiv\-:1105.3182 [astro-ph.CO]].

\bibitem{Hou:2011ec}
  Z.~Hou, R.~Keisler, L.~Knox, M.~Millea and C.~Reichardt,
  arXiv:1104.2333 [astro-ph.CO].
  
\bibitem{Archidiacono:2011gq} 
  M.~Archidiacono, E.~Calabrese and A.~Melchiorri,
  Phys.\ Rev.\ D {\bf 84}, 123008 (2011)
  [arXiv:1109.2767 [astro-ph.CO]].

\bibitem{zahn}
  T.~L.~Smith, S.~Das and O.~Zahn,
  Phys.\ Rev.\ D {\bf 85}, 023001 (2012)
  [arXiv:1105.3246 [astro-ph.CO]].

\bibitem{Hamann:2011hu} 
  J.~Hamann,
  JCAP {\bf 1203}, 021 (2012)
  [arXiv:1110.4271 [astro-ph.CO]].

\bibitem{Mangano:2005cc}
  G.~Mangano, G.~Miele, S.~Pastor, T.~Pinto, O.~Pisanti and P.~D.~Serpico,
  Nucl.\ Phys.\ B {\bf 729} (2005) 221
  [hep-ph/0506164].

\bibitem{calaratra}
  E.~Calabrese, M.~Archidiacono, A.~Melchiorri and B.~Ratra,
  arXiv:1205.6753 [astro-ph.CO].

\bibitem{Kopp:2011qd}
  J.~Kopp, M.~Maltoni and T.~Schwetz,
  arXiv:1103.4570 [hep-ph].

\bibitem{Giunti:2011gz} 
  C.~Giunti and M.~Laveder,
  Phys.\ Rev.\ D {\bf 84}, 073008 (2011)
  [arXiv:1107.1452 [hep-ph]].
  
\bibitem{Giunti:2011hn} 
  C.~Giunti and M.~Laveder,
  Phys.\ Rev.\ D {\bf 84}, 093006 (2011)
  [arXiv:1109.4033 [hep-ph]].

\bibitem{Giunti:2011cp} 
  C.~Giunti and M.~Laveder,
  Phys.\ Lett.\ B {\bf 706}, 200 (2011)
  [arXiv:1111.1069 [hep-ph]].

\bibitem{Karagiorgi:2012kw} 
  G.~Karagiorgi, M.~H.~Shaevitz and J.~M.~Conrad,
  arXiv:1202.1024 [hep-ph].

\bibitem{Donini:2012tt} 
  A.~Donini, P.~Hernandez, J.~Lopez-Pavon, M.~Maltoni and T.~Schwetz,
  arXiv:1205.5230 [hep-ph].

\bibitem{AguilarArevalo:2008rc}
  A.~A.~Aguilar-Arevalo {\it et al.}  [MiniBooNE Collaboration],
  Phys.\ Rev.\ Lett.\  {\bf 102}, 101802 (2009)
  [arXiv:0812.2243 [hep-ex]].

\bibitem{AguilarArevalo:2010wv} 
  A.~A.~Aguilar-Arevalo {\it et al.}  [The MiniBooNE Collaboration],
  Phys.\ Rev.\ Lett.\  {\bf 105}, 181801 (2010)
  [arXiv:1007.1150 [hep-ex]].

\bibitem{Zimmerman:2011hy} 
  E.~D.~Zimmerman [MiniBooNE Collaboration],
  arXiv:1111.1375 [hep-ex].
  
\bibitem{Djurcic:2012jf} 
  Z.~Djurcic [MiniBooNE Collaboration],
  arXiv:1201.1519 [hep-ex].

\bibitem{Abazajian:2012ys}
  K.~N.~Abazajian {\it et al.},
  arXiv:1204.5379 [hep-ph].

\bibitem{Mueller:2011nm}
  T.~.A.~Mueller, D.~Lhuillier, M.~Fallot, A.~Letourneau, S.~Cormon, M.~Fechner, L.~Giot and T.~Lasserre {\it et al.},
  Phys.\ Rev.\ C {\bf 83} (2011) 054615
  [arXiv:1101.2663 [hep-ex]].

\bibitem{Mention:2011rk}
  G.~Mention, M.~Fechner, T.~.Lasserre, T.~.A.~Mueller, D.~Lhuillier, M.~Cribier and A.~Letourneau,
  Phys.\ Rev.\ D {\bf 83} (2011) 073006
  [arXiv:1101.2755 [hep-ex]].

\bibitem{Huber:2011wv}
  P.~Huber,
  Phys.\ Rev.\ C {\bf 84} (2011) 024617
   [Erratum-ibid.\ C {\bf 85} (2012) 029901]
  [arXiv:1106.0687 [hep-ph]].

\bibitem{Giunti:2009xz}
  C.~Giunti and Y.~F.~Li,
  Phys.\ Rev.\ D {\bf 80} (2009) 113007
  [arXiv:0910.5856 [hep-ph]].

\bibitem{Palazzo:2011rj}
  A.~Palazzo,
  Phys.\ Rev.\ D {\bf 83} (2011) 113013
  [arXiv:1105.1705 [hep-ph]].

\bibitem{Palazzo:2012yf}
  A.~Palazzo,
  Phys.\ Rev.\ D {\bf 85} (2012) 077301
  [arXiv:1201.4280 [hep-ph]].

\bibitem{An:2012eh}
  F.~P.~An {\it et al.}  [DAYA-BAY Collaboration],
  Phys.\ Rev.\ Lett.\  {\bf 108} (2012) 171803
  [arXiv:1203.1669 [hep-ex]].

\bibitem{Ahn:2012nd}
  J.~K.~Ahn {\it et al.}  [RENO Collaboration],
  Phys.\ Rev.\ Lett.\  {\bf 108} (2012) 191802
  [arXiv:1204.0626 [hep-ex]].

\bibitem{Giunti-NUTURN-2012}
  C.~Giunti,
  Talk presented at nuTURN2012, 8-10 May 2012, LNGS, Assergi, Italy.


\bibitem{red}
 B.~A.~Reid {\it et al.},
  Mon.\ Not.\ Roy.\ Astron.\ Soc.\  {\bf 404} (2010) 60
  [arXiv:0907.1659 [astro-ph.CO]].

\bibitem{hst}
  A.~G.~Riess, L.~Macri, S.~Casertano, H.~Lampeitl, H.~C.~Ferguson, A.~V.~Filippenko, S.~W.~Jha, W.~Li {\it et al.},
  Astrophys.\ J.\  {\bf 730 } (2011)  119.
  [arXiv:1103.2976 [astro-ph.CO]].

\bibitem{Lewis:2002ah}
A. Lewis and S. Bridle,
Phys.\ Rev.\ D {\bf 66}, 103511 (2002) (Available from
\texttt{http://cosmologist.info}.)

\bibitem{Dodelson:2005tp}
  S.~Dodelson, A.~Melchiorri and A.~Slosar,
  Phys.\ Rev.\ Lett.\  {\bf 97}, 041301 (2006)
  [arXiv:astro-ph/0511500].

\bibitem{Hamann:2010bk}
  J.~Hamann, S.~Hannestad, G.~G.~Raffelt, I.~Tamborra and Y.~Y.~Y.~Wong,
  Phys.\ Rev.\ Lett.\  {\bf 105}, 181301 (2010)
  [arXiv:1006.5276 [hep-ph]].

\bibitem{Giusarma:2011ex} 
  E.~Giusarma, M.~Corsi, M.~Archidiacono, R.~de Putter, A.~Melchiorri, O.~Mena and S.~Pandolfi,
 Phys.\ Rev.\ D {\bf 83}, 115023 (2011)
 [arXiv:1102.4774 [astro-ph.CO]].

\bibitem{Archidiacono:2012gv} 
  M.~Archidiacono, E.~Giusarma, A.~Melchiorri and O.~Mena,
  arXiv:1206.0109 [astro-ph.CO].

\bibitem{Hannestad:2012ky}
  S.~Hannestad, I.~Tamborra and T.~Tram,
  arXiv:1204.5861 [astro-ph.CO].

\bibitem{Mirizzi:2012we} 
  A.~Mirizzi, N.~Saviano, G.~Miele and P.~D.~Serpico,
  arXiv:1206.1046 [hep-ph].

\bibitem{GonzalezGarcia:2010un} 
  M.~C.~Gonzalez-Garcia, M.~Maltoni and J.~Salvado,
  JHEP {\bf 1008}, 117 (2010)
  [arXiv:1006.3795 [hep-ph]].

\bibitem{Hamann:2011ge} 
  J.~Hamann, S.~Hannestad, G.~G.~Raffelt and Y.~Y.~Y.~Wong,
  JCAP {\bf 1109}, 034 (2011)
  [arXiv:1108.4136 [astro-ph.CO]].

\bibitem{Slosar:2006xb}
  A.~Slosar,
  Phys.\ Rev.\  D {\bf 73}, 123501 (2006)
  [arXiv:astro-ph/0602133].




\end{thebibliography}
\end{document}